# Micro-orchestration of RAN functions accelerated in FPGA SoC devices

Nikolaos Bartzoudis[1], José Rubio Fernández[1], David López-Bueno[1], Godfrey Kibalya[2] and Angelos Antonopoulos[2]

[1]Centre Tecnològic de Telecomunicacions de Catalunya (CTTC-CERCA), Castelldefels, Barcelona, Spain
{nbartzoudis, prubio, dlopez}@cttc.es

[2]Nearby Computing S.L., Barcelona, Spain
{godfrey.kibalya, aantonopoulos}@nearbycomputing.com

*Abstract*—This work provides a vision on how to tackle the underutilization of compute resources in FPGA SoC devices used across 5G and edge computing infrastructures. A first step towards this end is the implementation of a resource management layer able to migrate and scale functions in such devices, based on context events. This layer sets the basis to design a hierarchical data-driven micro-orchestrator in charge of providing the lifecycle management of functions in FPGA SoC devices. In the O-RAN context, the micro-orchestrator is foreseen to take the form of an xApp/rApp tandem trained with RAN traffic and context data.

## I. Introduction

THE orchestration of cloud computing resources traditionally focuses on how workloads could be scaled, migrated, and executed across computing clusters that feature multicore processors and graphics processing units (GPU). With the shift of fifth generation (5G) services towards the telco edge, new considerations have arisen. While the far edge compute infrastructure resembles to small-scale datacenters with homogeneous processing resources, the near and extreme edge typically feature a heterogenous ecosystem of computing elements that span from small footprint GPUs, embedded processors, low-power micro-controllers, field programmable gate arrays (FPGA) and application-specific integrated circuits (ASIC), up to System-on-Chip (SoC) devices that combine the previous computing elements in the same silicon fabric [1]. Another striking difference of the far/extreme edge as opposed to near edge and central clouds, is that processing resources are notably scarce, highly distributed and need to be managed efficiently [2] by exploiting at maximum their compute capacity, while satisfying stringent energy and latency requirements.

In such edge environments, the functional operation of hierarchical orchestrators needs to be extended by engaging smaller-scale micro-orchestrators [3] able to efficiently leverage compute resources and address the heterogeneity of edge computing elements. This is especially challenging in complex SoC devices that embed general-purpose and accelerator-driven processing elements [4]. The fine grain micro-orchestration of the compute resources in complex SoC accelerators is also crucial to satisfy agile, timely and energy efficient resource management. Fig. 1 provides an overview of resource orchestration scales across the compute continuum w.r.t control loop timescales and accelerated workloads.

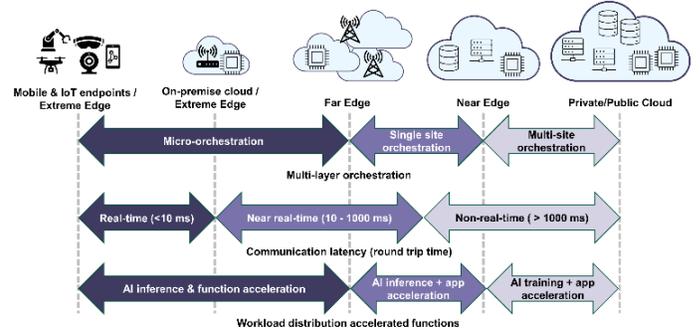

*Fig. 1:The different scales of resource orchestration.*

## II. FPGA SoC Devices In The 5G Compute Continuum

SoC devices with FPGA resources are used as function accelerators across the 5G radio access network (RAN) and cloud infrastructures. Having as a reference the open RAN Alliance (O-RAN) architecture, FPGA SoC devices are encountered i) in open radio units (O-RU) accelerating low physical layer (PHY) digital signal processing (DSP) functions, ii) in network interface cards (NIC) implementing the fronthaul interface, iii) in open distributed units (O-DU) accelerating specific high-PHY functions (i.e., channel coding), and v) in the near-real-time RAN intelligent controller (RIC) hosting the inference of different machine learning (ML) models as extended applications (xApp). On top of that, FPGA devices are used for accelerating applications in different scales of cloud computing infrastructures.

The underlying complex architecture of these multi-processing devices and the heterogeneity of the embedded processing elements, makes challenging and cumbersome the combined virtualization of the underlying compute resources, the exposure of deep telemetry data and, consequently, the deployment of such devices in Kubernetes clusters with full resource observability. Equally challenging is the run-time fine grain adaptive management of the computing resources either at task or at function level. Different efforts both from the industry and academia have been trying to address these challenges, focusing on concrete use cases and offering solutions tailored for specific families of SoC devices. For instance, Microsoft's Catapult v2 [5] work focuses on the offloading of network processing from the embedded processor to the FPGA area of the SoC device over Microsoft's Azure framework. Another work in IBM Research divides the FPGA spatially into distinct application regions, where hardware accelerated applications are to be programmed; the Service Logic secures access to shared off-chip memory and a dedicated host Processor-based server [6]. Amazon's AWS F1 instance offers connectivity to eight FPGA cards which are connected to a single physical server and a dedicated FPGA-only interconnection network [7]. Multiple academic works have also explored the deployment of FPGAs in cloud environments, but their thorough review goes beyond the scope of this paper.

In current commercial deployments, FPGA SoC devices are typically used as monolithic compute resources. This means that a single function reserves the entire FPGA area per timeslot (i.e., multi-tenancy is not applied), or multiple functions from different users reserve the entire FPGA area on a permanent basis (i.e., no time division multiplexing of resources is applied). The remaining compute resources of the SoC device are underexploited in the spatial and time domain. This deficiency becomes highly critical in edge environments due to the scarcity of computing resources, the elevated processing requirements of 5G and beyond edge applications and the battery-limited operation of terminal devices. This is precisely why edge infrastructure owners need to flexibly leverage the full capacity of such devices in a fine grain mode.

### III. CONTRIBUTION AND DEVELOPMENT ROADMAP

The main contribution of this work is to provide a resource management layer for functions running in FPGA SoC devices [1], which along with the run-time reconfiguration framework presented in [4] (i.e., joint management of interdepended software and FPGA functions) form the necessary substrate for designing an intelligent closed-loop micro-orchestrator. The latter is currently under development, and will be able to reconfigure, scale, migrate, or replace functions across the SoC fabric based on different intelligent control loops. The micro-orchestrator could be seen as a hierarchical data-driven intelligent controller that will be built by training an artificial intelligence (AI) model with RAN traffic data, context-related information and the on-chip telemetry data (e.g., execution time, power consumption, throughput of embedded buses). Using the notions of O-RAN, the micro-orchestrator will either take the form of a combined rApp and xApp, or a real-time application located at the extreme edge [8]. The micro-orchestration will target the accelerated functions residing in O-RUs and O-DUs, whose rather static operation is expected to be challenged in 6G use cases targeting real-time control loops [9].

As part of this ongoing and future roadmap, we present in this paper a FPGA SoC system able to reconfigure its underlying functions based on events that are detected by a computer vision edge application. This context-driven function reconfiguration will be integrated with the low physical-layer (low-PHY) of a commercial O-RU featuring the functional split 7.2. The O-RU has already been interfaced with the O-RAN stack and 5G core of the open-air interface (OAI). The mentioned integration is also work in progress and will be validated in a smart city use case (i.e., autonomous tram edge-assisted services) in the context of the project VERGE [3].

### IV. FPGA SoC RESOURCE MANAGEMENT LAYER

The computer vision application is hosted in the AMD Kria KV260 Vision AI Starter Kit [10] and the function that is used for reconfiguration purposes in the AMD Zynq UltraScale+ RFSoC ZCU111 Evaluation Kit [11], thereafter denoted as RU emulation platform. The selected reconfigurable function is a fast Fourier transform (FFT) processing block. This is either hosted in the ARM A53 processor of the ZCU111 radio frequency SoC (RFSoC) device using the open source FFTW implementation [12], or in the programmable logic (PL) area of the same device using the AMD FFT LogiCore [13]. The computer vision application in the KV260 edge node detects events, which are then communicated to the ZCU111 platform. To do so, a software hook has been added to the KV260 to count and expose events to the ZCU111 platform using a socket network connection.

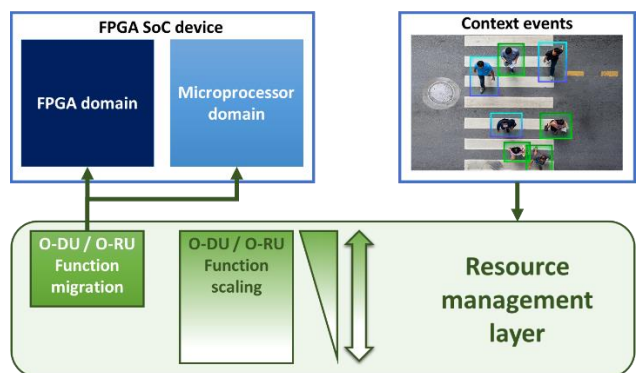

*Fig. 2: Function reconfiguration in FPGA SoC devices.*

The FPGA SoC resource management layer is a Linux process running in the ARM A53 processor of the RFSoC device able to seamlessly apply: i) function migration, ii) function scaling, iii) function placement, and iv) function reconfiguration. The implementation of the last two options was presented in [4], and thus is not covered in this paper. As seen in Fig. 2, this work focuses on the run-time function scaling and migration from a software to a hardware-accelerated execution domain, based on the events detected by the edge application. As it is shown in Fig. 3, a web camera is connected to the KV260 edge node, where a computer vision application running in the Zynq UltraScale+ device detects faces. In the following, we describe the different hardware, firmware and software components comprising the two main platforms.

#### A. Edge node

- **PL functions**: The PL part of the Zynq UltraScale+ device performs the processing of the video signal. Furthermore, it hosts a natural language processing computer vision application [14], which implements face detection using a deep learning processor (DPU). The latter features the pre-compiled DenseBox face detection model from the Xilinx Vitis-AI Model Zoo (i.e., Network model: cf densebox wider 360 640 1.11G 1.2). The output video signal is overlayed with a frame surrounding each detected face and it is constantly updated in the 2D video space domain. This output signal is displayed on a monitor.

- **Application processing unit (APU) functions**: A Linux application hosted in the APU configures and initializes the PL part. On top of that, a module was added to process the DPU output and count the number of detected faces. Upon event occurrence, the number of detected faces is notified to the ZCU111 RFSoC device. To do so, a socket network connection is created between the KV260 and ZCU111 boards and a message is sent by an APU application.

*B. RU emulation platform*

- **PL functions:** The PL accelerated FFT [13] uses a signal located in the platform's DDR memory. The output of the FFT is stored in another area of the DDR memory. The FFT processing block is always configured in the PL area, but when not in use, it is deactivated through clock-gating signaling, to reduce the PL dynamic power consumption.
- **APU functions:** A socket client application receives the messages from the KV260 board and retrieves the issued events (number of faces detected). The APU executable includes a precompiled FFTW. Switching between this FFT software version and the FPGA-accelerated PL FFT version is made feasible by a reconfiguration controller (i.e., part of the FPGA SoC resource management layer), which performs FFT function migration plus scaling at run-time (i.e., variation of the number of points of the FFT). The reconfiguration controller takes the following actions upon event detection: i) if 0 faces are detected then the FFTW is used (8 points FFT), ii) if 1 face is detected the FFTW is used (1024 points FFT), iii) if 2 faces are detected the FFT LogiCore is used in PL (2048 points FFT), iv) if more than 2 faces are detected the FFT LogiCore is used in PL (4096 points FFT). A performance comparison is also applied by calculating the mean squared error (console window in Fig. 3) between the software-executed FFTW function (floating point operations) and the FPGA accelerated FFT LogiCore function (fixed-point precision).
- **Power monitoring:** An application was created to monitor the power consumption of the ZCU111 RFSoC device. The embedded Linux system running in the APU periodically reads the monitoring data from the on-chip voltage sensors and dumps into the Linux file system. The power monitoring application sends the metrics to a host machine through a socket network connection, where they are visualized in a Python application (middle left in Fig. 3).

## V. CONCLUSIONS

Apart from the FFT, other RAN functions (e.g., channel coding) or edge applications could be tested with the framework presented in this paper. The FPGA SoC resource management layer is currently extended to include the real-time ARM R5 processor and the NEON instructions of the ARM A53 processor as function migration processing options. Also, the experimental setup is benchmarked under different operating scenarios. Finaly, the reconfiguration framework of [4] will be integrated in the resource management layer. According to the boot time estimation tool [15] a partial PL reconfiguration bitstream occupying for instance 5% of the ZCU111 flash memory, would approximately require 10 ms to be transferred in a PL reconfigurable region. Thus, faster reconfiguration strategies like the activation/deactivation of PL functions, or longer-term FPGA SoC resource usage forecasting shall be contemplated in real-life use cases. Such scenarios will be validated once the FPGA SoC resource management layer is integrated in the low-PHY of a commercial O-RU.

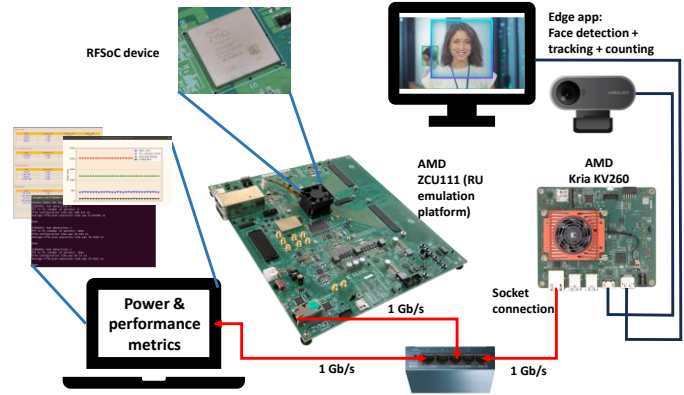

*Fig. 3: Hardware diagram of the experimental setup.*

ACKNOWLEDGMENT

This work was supported in part by the Horizon Europe SNS JU VERGE project funded by the European Commission (ID 101096034), the project ORIGIN (PID2020-113832RB-C22) funded by MICIN (Gobierno de España), the projects FREE6G-RegEdge (TSI-063000-2021-144) and 6GBLUR-Smart (TSI-063000-2021-56) funded by MINECO (Gobierno de España), and the grant 2021 SGR 00772 funded by AGAUR (Generalitat de Catalunya).